\documentstyle[11pt,aaspp4,tighten]{article}

\begin{document}

\font\twelvei = cmmi10 scaled\magstep1 
       \font\teni = cmmi10 \font\seveni = cmmi7
\font\mbf = cmmib10 scaled\magstep1
       \font\mbfs = cmmib10 \font\mbfss = cmmib10 scaled 833
\font\msybf = cmbsy10 scaled\magstep1
       \font\msybfs = cmbsy10 \font\msybfss = cmbsy10 scaled 833
\textfont1 = \twelvei
       \scriptfont1 = \twelvei \scriptscriptfont1 = \teni
       \def\mit{\fam1 }
\textfont9 = \mbf
       \scriptfont9 = \mbfs \scriptscriptfont9 = \mbfss
       \def\bmit{\fam9 }
\textfont10 = \msybf
       \scriptfont10 = \msybfs \scriptscriptfont10 = \msybfss
       \def\bmsy{\fam10 }

\def\etal{{\it et al.~}}
\def\eg{{\it e.g.}}
\def\ie{{\it i.e.}}
\def\lsim{\raise0.3ex\hbox{$<$}\kern-0.75em{\lower0.65ex\hbox{$\sim$}}} 
\def\gsim{\raise0.3ex\hbox{$>$}\kern-0.75em{\lower0.65ex\hbox{$\sim$}}} 
 
\title{Spectral Properties of Accretion Disks Around Black Holes II -- 
Sub-Keplerian Flows With and Without Shocks\footnote[1]
   {Submitted:11th Nov. 1996; Accepted: Jan. 22 1997; Astrophysical Journal (in press)}}
 
\author{Sandip K. Chakrabarti\footnote[2]
{On leave from TIFR, Mumbai 400005, India}}
\affil{S.N. Bose National Center for Basic Sciences
JD Block, Sector III, Salt Lake, Calcutta 700091, India\\
and\\
Tata Institute of Fundamental Research, Mumbai, 400005, India\\
e-mail: chakraba@tifrc2.tifr.res.in}

\vskip 1cm
\begin{abstract}

Close to a black hole, the density of the sub-Keplerian accreting matter 
becomes higher compared to a spherical flow
due to the presence of a centrifugal barrier
independent of whether or not a standing shock actually forms. This hot dense 
flow intercepts soft photons from a cold Keplerian disk and reprocesses them 
to form high energy X-rays and gamma rays. We study the spectral 
properties of various models of accretion disks where a Keplerian disk 
on the equatorial plane may or may not be flanked by a sub-Keplerian 
disk and the sub-Keplerian flow may or may not possess standing shocks. 
From  comparison with the spectra, we believe that the observed 
properties could be explained better when both the components (Keplerian 
and sub-Keplerian) are simultaneously present close to a black hole,
even though the sub-Keplerian halo component may have been produced out of the
Keplerian disk itself at larger radii.
We are able to understand soft and hard states of black hole candidates, 
properties of X-ray novae outbursts, and quasi-periodic oscillations of 
black hole candidates using these two component models. We fit 
spectra of X-ray novae GS1124-68 and GS2000+25 and satisfactorily 
reproduce the light curves of these objects. 

\end{abstract}


\clearpage
 
\section{INTRODUCTION}
 
Recently, Chakrabarti \& Titarchuk (1995, hereafter referred 
to as Paper I) proposed that in order to understand the soft and hard 
states of black hole candidates, one need not look for any elusive 
`Compton cloud' or any so-called magnetized corona. One can 
satisfactorily explain all the major quasi-steady properties of
black hole candidates by a two component accretion flow model (TCAFM).
Based on analytical solutions of generalized accretion disk models 
(Chakrabarti, 1990; Chakrabarti \& Molteni, 1995; Chakrabarti 1996a, hereafter 
C96a) they showed that an accretion disk should naturally segregate into two
regions, one a Keplerian disk on the equatorial plane and the 
other a sub-Keplerian halo flanking the Keplerian disk, although eventually
the Keplerian component should also become sub-Keplerian close to the black
hole in order to satisfy boundary conditions on the horizon. 
The sub-Keplerian halo is hot, optically thin, and faces a centrifugal barrier 
close to the black hole; as a result, its density increases much faster 
than that of a spherical Bondi flow. This `puffed up' optically thin 
gas located just outside the horizon may intercept soft photons coming from
the cooler Keplerian disk and re-emit them as photons of higher energy (hard 
X-rays and gamma-rays in the case of galactic black hole candidates, and UV 
and soft X-rays in the case of massive black hole candidates). Detailed study 
of the behavior of the sub-Keplerian component (Chakrabarti 1989, hereafter 
C89; C96a) shows that in a considerable region of the parameter 
space, the flow actually forms a standing shock where the density and 
temperature rise abruptly. In the case of the formation of
{\it strong} shocks (where density jumps by a factor of four) the
electron number density is high in the post-shock region, and Paper 
I shows that the accretion rate of the Keplerian disk where a hard state 
(typically, when the energy spectral index $\alpha \sim 0.5-0.8$) to a soft 
state (typically, when the energy spectral index $\alpha \sim 1.3-1.8$) 
transition takes place is around $0.1-0.3{\dot M}_{Edd}$ (where 
${\dot M}_{Edd}$ is the Eddington rate), depending on the accretion rate of 
the sub-Keplerian component. In the future, we shall call this model 
as TCAFM1. 

However, in the general classification of C89 and Chakrabarti
(1996b, hereafter C96b), the rest of the parameter space does not allow the
formation of standing shocks. The density and temperature enhancements are 
gradual, though ultimately achieving values close to the hole similar
to those of the post-shock flows (C96a and Chakrabarti, 1997 hereafter C97). 
We discuss the nature of these regions below as well. Existence of these 
solutions enables us to explore other possible models which we carry out in 
the present paper. These models are: (a) TCAFM2 -- Here 
the shock is weak or absent altogether, but the flow feels a centrifugal 
barrier due to the angular momentum. (b) TCAFM3 -- Here the 
sub-Keplerian component does not have any angular momentum, and is basically 
a spherical inflow. This case is considered to bring out the important fact 
that the observed properties of black hole candidates may definitely require
some angular momentum. And finally, (c) SCAFM, Single Component 
Accretion Flow Model -- Here the sub-Keplerian halo is 
completely non-existent, and only the Keplerian component 
becomes sub-Keplerian close to the black hole.  It is to be noted that
another single component (inviscid) solution is possible where the flow 
remains fully sub-Keplerian and the post-shock flow (with or without shocks)
intercepts soft-photons from the pre-shock region. This has been
studied by Chakrabarti \& Wiita (1992) in the context of
spectra of active galaxies and quasars.

In this context we differentiate two types of black holes candidates 
depending on the time scales in which their states are changed. If the 
candidate is an X-ray nova, then  it may be repeated in a very long time 
scale (several tens of years, say). This time scale is probably dictated 
by the limit cycle behavior (Cannizzo, 1993) in which the viscosity at the 
outer edge of the Keplerian disk rises and falls periodically and the mass
accretion rates also can change. In the quiescent states, with low accretion 
rate, the hot flow accretes basically as a sub-Keplerian disk since the 
deviation from a Keplerian disk takes place very far away from a black hole 
($X_K >> 1$, Paper I, C96). During the rising phase of a nova outburst, the
hard X-ray brightens in a matter of a few ($2 - 5$) days which may correspond 
to the viscous (upper branch of the limit cycle) time scale in which 
Keplerian disk component rapidly comes closer to the black 
hole. Keeping this in mind, we also study a variation of TCAFM2 and 
TCAFM3 models (called TCAFM2$_2$ and TCAFM3$_2$ respectively) in which the 
transition radius $X_K$ is varied in a viscous time scale. In the decaying 
phase of a nova, both the accretion rates as well as $X_K$ (where deviation 
from a Keplerian  disk takes place) are expected to change in an as yet poorly 
understood manner. Nevertheless, the soft to hard transition could be 
understood generally in terms of  the relative abundance of Keplerian
and non-Keplerian matter, as we show below. In the case of otherwise inactive 
black holes (e.g., away from novae activities), the soft state to hard state 
(and vice versa) may take place several times in a matter of days or months. 
These cases could be due to variation of viscosity in these time scales 
and the Keplerian and sub-Keplerian components could be re-distributing 
their rates accordingly, while maintaining the sum of the rates roughly 
constant. Keeping these systems in mind, we study a model (TCAFM1$_2$) 
where the sum of the rates are kept fixed.

The current Paper is organized as follows: In the next Section, we classify 
all the solutions in thin accretion flows. Using this as the reference point, 
we discuss our disk models and the way the spectra are computed. In \S 3, we 
present solutions of various models. In \S 4, we compare solutions of our 
models with observations of X-ray novae. Finally, in \S 5, we present 
our conclusions.

\section{DISK MODELS}

In Fig. 1, we classify the {\it entire} parameter space according to the 
type of solutions of thin inviscid flow that are prevalent around a 
Schwarzschild black hole. Here we use two conserved 
parameters, namely, specific energy 
${\cal E}$ and specific angular momentum $l$ for classification (as opposed
to, say, entropy accretion rate ${\dot{\cal M}}$ and $l$ as used in 1.5D flow 
in C89). The adiabatic index $\gamma=4/3$ has been chosen. The whole space is 
divided into nine regions marked by $N$, $O$, $NSA$, $SA$, $SW$, $NSW$, $I$,
$O^*$, $I^*$. The horizontal line at ${\cal E}=0$ corresponds to the rest
mass of the flow. 

The parameters from region $N$ do not produce any transonic solution.
The solutions from the region `O', which have very low angular momentum 
and energy are similar to a Bondi flow and have only the outer sonic point.
The solutions from the regions $NSA$ and $SA$ have two `X' type sonic points
with the entropy density $S_o$ at the outer sonic point {\it less} than the
entropy density $S_i$ at the inner sonic point. However, flows from $SA$
pass through a standing shock since the Rankine-Hugoniot
condition is satisfied. The entropy generated at the shock,
$S_i-S_o$, is advected towards the black hole to enable the flow to pass
through the inner sonic point. Rankine-Hugoniot condition is not satisfied
for flows around a black hole from the region $NSA$. However, for flows around
a neutron star the shock condition is satisfied right outside the star surface 
(C89). Numerical simulations show (Ryu, Chakrabarti \& Molteni, 1997) that the 
flow from this region could be very unstable and could exhibit periodic 
changes in emission properties as the flow constantly tries to form the shock 
wave, but fails to do so. These solutions explain quasiperiodic oscillations 
very satisfactorily. 

The solutions
from the regions $SW$ and $NSW$ are very similar to those from $SA$ and $NSA$. 
However, $S_o \geq S_i$ in these cases. Shocks can form only 
in winds from the region $SW$. The shock condition is not
satisfied in winds from the region $NSW$ even though two
sonic points are present. This may make the $NSW$ flow
unstable as well. A flow from region $I$ only has the inner sonic
point and thus can form shocks (which require
the presence of two saddle type sonic points) only if the inflow is already 
supersonic due to some other physical processes (such as flaring of the
inflow, or capturing of companion winds; see, C90, C97). Each solution 
from regions $I^*$ and $O^*$ has two sonic points (one `X' and one `O')
only but none produces any complete and global solution. The region $I^*$
has an inner sonic point but the solution does not extend subsonically
to a large distance. The region $O^*$ has an outer sonic point, but the
solution does not extend supersonically to the horizon.
When a significant viscosity is added, the closed topologies of $I^*$ and $O^*$
open up as described in C90ab and C96, and the flow may join with a cool
Keplerian disk with ${\cal E} <0$. These special solutions of viscous 
transonic flows need not have centrifugally supported shock waves as they have
only one inner sonic point. However, hot flows deviating from a Keplerian 
disk, or sub-Keplerian winds from companions, or cool flows subsequently 
energized by magnetic flares (for instance) will have ${\cal E}>0$, and thus 
could have standing or periodically varying shock waves as discussed above. 
The post-shock region or the enhanced density region radiates most of 
the observed hard radiation. The present classification is done using 
thin flows in the Paczy\'nski--Wiita
(1980) potential. A similar division of parameter space in the Kerr geometry 
is presented elsewhere (Chakrabarti, 1996c).

In Paper I, we have already discussed in detail the nature of the 
two component hydrodynamical model. In the present paper,
in view of the classification described above,
we not only study that model, we also assume that the shock
need not be always strong as was assumed in Paper I, or need not always be
present. Fig. 2 shows schematically the general nature of the
multi-component accretion flows. If the basic assumption that the
viscosity falls off with height is correct, then when the accretion rate
in the sub-Keplerian component is low, then the lowest viscosity 
region separates out (at $X_{K2}$) as a giant big thick disk 
(of size $\sim 10^4 R_g$) while the intermediate viscosity region separates 
out (at $X_{K1}$) as a sub-Keplerian flow which may or may not have a 
shock wave (at $X_{S}$). If the shock does not form then $X_S \sim X_{K1}$
and $X<X_{S}$ is simply the centrifugal barrier supported dense region 
which reprocesses the soft photons in the same way as the post-shock
region of Paper I. In the hard state, the giant (geometrically) thick disk
(which is like an extended atmosphere of optical depth of the order unity) 
reprocesses the hard and soft radiations and may even remove the
soft bump altogether (as seen in Cyg X-1, see, 
Kazanas, Hua \& Titarchuk, 1997). When the viscosity is increased and 
the Keplerian accretion rate is increased, the giant thick disk also
collapses (this can be understood from the Fig. 10 of Paper I) 
thereby exposing the Keplerian disk along with the cool post-shock region
to the observer.

In the case when the angular momentum 
of the sub-Keplerian component is negligible (TCAFM3), the Keplerian 
disk is simply assumed to be `embedded' in a wedge-shaped 
infalling `Bondi' cloud which changes its 
conical thickness at the place where the Keplerian
disk also becomes sub-Keplerian. At this place, the thickness (which 
determines the fraction of soft-photons intercepted by the cloud) 
is computed from the average local temperature of the gas, which is 
roughly the same as that when the shock is present. The actual
thickness and interception may depend on details, but the 
conclusion from this model does not. In the case where shocks are not present 
or are weak (TCAFM2), the density is increased due to the centrifugal barrier 
as  matter comes closer to the black hole. The detailed solution of the 
governing equations  (e.g., Fig. 7a of C96a, and Fig. 3 of C97) shows that 
independent of whether a shock is present or not, 
for a given angular momentum at the inner edge of 
the disk, the flow has roughly the same distribution of velocity
and density, though in the presence of shocks this variation is more abrupt. 
In the Single Component Accretion Flow Model (SCAFM), the sub-Keplerian 
halo component is not assumed to be present
at all. The Keplerian disk becomes sub-Keplerian at 
$X_K$, and the density thereon was computed using the standard procedure
with the angular momentum barrier taken into account as in TCAFM2. As will be 
shown below, this case always tends to produce soft states.

In the models TCAFM2$_2$ and TCAFM3$_2$, the location $X_K$ is varied and the
corresponding spectra are computed. Generally speaking, in the presence 
of a steady sub-Keplerian component, the object should go from a hard state 
to a soft state as $X_K$ becomes smaller (due to variation of viscosity for 
example, see, C90, Paper I, C96ab) and more and more soft photons are 
intercepted by the sub-Keplerian enhanced density region close to the
black hole. In the rising phase of novae outbursts such a process may 
indeed be taking place on viscous time scales dictated by the higher 
value of viscosity during the limit cycle, while in the decaying phase of 
novae outbursts such a process may take place in a reverse order, on time 
scales dictated by the smaller viscosity during this phase of the limit cycle.
This is suggested by the fact that viscosities in the two branches 
are off by almost two orders and 
the rise time scale of a novae outburst is also about two orders of 
magnitude shorter compared to the decaying time scale. Of course, 
the effect due to varying $X_K$ is superimposed on the (exponentially
or linearly, as the case may be) decaying mass accretion rates, as 
discussed in \S 4 below.

In all the model calculations,  we assume the mass of the black hole to be
$1M_\odot$. This is uncorrected for the spectral hardening (Shimura \& 
Takahara, 1995) factor $f$ as in Paper I. Thus, assuming  $f\sim 1.9$, 
our solutions here are valid for a black hole of mass $f^2 M_\odot \sim 
3.6 M_\odot$. The general results are valid even for supermassive black 
holes in a trivial way, since the electron temperature in the 
enhanced-density-region (which serves as the electron cloud in our model) 
is very insensitive to the black hole mass. (In fact the virial temperature 
is independent of the black hole mass, but considering 
corrections due to cooler disk photon temperatures, we
find empirically that the electron temperature varies as
$T_e \sim M_{BH}^{0.04-0.1}$, keeping all other non-dimensional inputs
same.). The spectra are computed assuming a line of sight angle of 
$\sim 37^{\circ}$ (cos $i=0.8$). The accretion rates are always
quoted in units of the Eddington rate, and the distances are measured in 
units of the Schwarzschild radius $R_g= 2GM_{BH}/c^2$.  In some of the
cases the effect of bulk motion Comptonization (Paper I and references therein;
Ebisawa, Titarchuk \& Chakrabarti 1996) has been added just for comparison.

\section{SPECTRAL PROPERTIES OF DIFFERENT ACCRETION FLOW MODELS}

To remind the readers, we describe the procedure of computation of
the spectra here, but details are in Paper I. We generally assume
two accretion rates, ${\dot m}_d$, the rate of the Keplerian component, and 
${\dot m}_h$, the rate of the sub-Keplerian halo component. It is to be noted 
that these two independent rates obviates the need to use 
one single rate (sum of Keplerian and sub-Keplerian) and 
an unknown viscosity parameter. 
However, depending on astrophysical circumstances, the sum of the 
two rates may or may not be constant with time (as discussed 
in the Introduction). Once ${\dot m}_d$ is
assumed, the zeroth order temperature distribution of the Keplerian disk is
computed from a standard Shakura-Sunyaev (1973) disk. Similarly, once the
${\dot m}_h$ is chosen, the density distribution is computed from  a
standard hydrodynamical model (Paper I) and the temperature is computed
using a two-temperature disk model with bremsstrahlung, Comptonization,
proton-electron Coulomb coupling, inverse bremsstrahlung, etc. The matter
from both components are mixed below $X_K$ ($=X_s$ if shocks are present).
The electron temperature in the sub-Keplerian inner region below $X_K$
is computed and then averaged using the prescription given in Paper I
and the spectral index is computed using standard procedures
(Sunyaev \& Titarchuk, 1985; Titarchuk 1994). The feedback from the
sub-Keplerian cloud onto the cooler Keplerian disk is computed using the
proper albedo (Paper I) and the spectral index is recomputed using the
modified cool disk temperature. The procedure is iterated until the spectral
index, the disk temperature distribution, and the electron 
temperature all converge.

\subsection{When a Strong Shock is Present}

In this case we study the model TCAFM1. Fig. 3 shows the nature of the 
spectra. Here, we choose ${\dot m}_h=1$ and ${\dot m}_d=1,\ 0.1,\ 0.01,
\ 0.001,\ 0.0001$ as marked. The shock location is chosen at $X_s=10$,
which is the most likely place in a Schwarzschild geometry (Paper 1).
The inner edge is chosen at $X_i=3$, but the spectra from thermal
Comptonization alone depend weakly on the exact location as the surface
area of this region goes down rapidly. In this model, the system goes from
hard state to the soft state when the accretion rate of the Keplerian
component is increased. The variation of the spectral index will be 
presented in \S 3.5. The dotted curve is drawn incorporating consideration 
of the bulk motion Comptonization (Paper I), 
which produces a weak hard tail even in the soft state.

Occasionally, the net accretion rate of the inflow may remain
constant and the internal variation of viscosity would simply re-distribute
Keplerian and sub-Keplerian components differently, as discussed in Paper I.
This is our TCAFM1$_2$ model. Fig. 4a depicts the spectral variation for one 
such case where the sum is kept at twice the Eddington accretion rate 
(${\dot m}_d+{\dot m}_h = 2$). The curves are drawn for various 
Keplerian disk rates as marked. In this case, the gravitational
potential energies of matter are released in soft  and
hard components at different rates, depending on the availability of soft
photons in the Keplerian disk. The net energy released is roughly constant,
as is observed in many black hole candidates (Zhang et al., 1997). The 
pivoting property is also seen to be present. The shock location is 
assumed to be $X_s=10$ here. By increasing $X_s$, the pivoting energy
could be reduced. (Typical variations of $X_s$ with angular momentum 
in a vertically averaged model of the flow are presented in C89.) 
The effect of the bulk motion Comptonization for the case 
${\dot m}_d=1$ is also shown in dotted curves. In Fig. 4b, we present 
the variation of the mean electron temperature
of the sub-Keplerian cloud close to the black hole with the disk accretion rate
(spectral index variation will be presented in \S 3.5).
Here, the electron temperature goes down as the Keplerian rate goes up.
The solid curve shows the variation keeping the halo rate fixed
(the case corresponding to Fig. 3 of TCAFM1) and the dashed curve shows 
the variation keeping the sum of the rates fixed (the case
corresponding to Fig. 4a of TCAFM1$_2$). We note that in the latter case, the 
mean electron temperature is generally constant (around $100-150$keV) 
in the hard state, while at about ${\dot m}_h\sim 0.3-1$, the temperature 
goes down to a few keV rather catastrophically and the object reaches
the soft state. In fact, given that the efficiency of extraction of
energy from the sub-Keplerian component is less than what is 
allowed from the conversion of the gravitational potential
energy, it is likely that the halo rate actually increases much faster
than what is dictated by merely keeping the sum constant. This is probably
supported by observed rigorous constancy of the spectral index in each state.

\subsection{When the Sub-Keplerian Component has no Angular Momentum}

This model is TCAFM3.  Figure 5 shows the spectral variation with the
Keplerian disk accretion rate ${\dot m}_d$ (as marked) while the halo rate
${\dot m}_h$ is chosen to be unity as before.
In this case, the absence of angular
momentum reduced the electron density of the sub-Keplerian Compton
cloud at the inner edge. Therefore, for a given halo rate, the
hard X-ray emission is lower as can be seen by comparing with 
Fig. 3. Furthermore, it takes fewer soft photons (namely, smaller
${\dot m}_d$) in order to bring the system to soft state, as is clear from 
the spectra. The actual comparison of the spectral indices is done in \S 3.5.
The effect of bulk motion Comptonization is similar as in the earlier models
and is not shown.

In this particular Model, it is instructive to study the 
variation of spectral index when the transition radius
$X_K$ (where the disk deviates from a Keplerian disk) is varied. 
This is our TCAFM3$_2$.
In presence of low angular momentum flow, viscosity influences the
distribution very strongly and $X_K$ decreases with the increase
of viscosity (for a given inflow angular momentum on the horizon). In Fig. 6a
the results are shown. As $X_K$ is monotonically decreased,
the black hole goes from hard to soft state. The halo rate 
is kept fixed at ${\dot m}_h=1$ and the Keplerian rate is
fixed at ${\dot m}_d=0.05$.  It is to be contrasted with 
the case when the sub-Keplerian halo is absent where the
disk always remains in a soft state (see below).  In Fig. 6b,
we show how the spectral index is changed as the transition radius $X_K$ 
is reduced. A possible variation of this model is obtained where the Keplerian
disk is originally situated very far away ($\sim 10^4$) as in a low viscosity, 
low accretion rate, disk (see, Fig. 2a of C96a) and then it 
approaches the black hole as viscosity is enhanced. In this case, the
object, though very faint, has the signature of a soft state. But as the 
viscosity starts increasing and $X_K$ starts getting smaller, 
the disk first becomes very bright in hard X-ray and then 
goes back to the soft state once more with fainter hard X-ray.
This type of behavior is common in novae outbursts (Ebisawa et al. 1994; 
Kitamoto et al. 1992). In Fig. 7a-b, we 
show the spectral variation in this case. Values of $X_K$ are marked
on the curves. We fix ${\dot m}_h=1$ for illustration purposes.
In Fig. 7a, we choose ${\dot m}_d=0.001$, and in Fig. 7b, we choose
${\dot m}_d=0.01$. In reality, both rates should increase 
with time when the rising phase of an X-ray outburst
is considered. The spectral index variation
in this case will be shown in \S 3.4. Note that in Fig. 7a, the 
spectra remains basically in hard states, since the intercepted 
photon number is too small compared to the electron density. In Fig. 7b,
the Keplerian rate is $10$ times higher, and as a result, the flow
goes to the soft state as $X_K$ decreases to around $X_K\sim 30$.

\subsection{When the Shock is Weak or Absent}

Here, angular momentum is present but the shock conditions are not satisfied. 
This is our model TCAFM2. Fig. 8 shows the variation of the spectral index 
with $X_K=10$ and ${\dot m}_h=1$. The hard to soft state transitions 
take place due to an increase in the number of soft photons relative to the 
electron numbers as before;  the soft photon numbers are characterized 
by the disk accretion rates ${\dot m}_d$ indicated on the curves. 
The result is very similar to TCAFM1, where a strong shock is present.
This is because the density variation in the post-shock region is similar
to the density variation in the centrifugal barrier for the same
value of angular momentum (which is chosen to be $1.837$, the
marginally stable value, in the present run). The variation of the
spectral index with ${\dot m}_d$ is shown in \S 3.5.

As in the previous sub-section, it may be instructive to study the spectral 
properties when the transition radius $X_K$ is varied. This corresponds to 
TCAFM2$_2$ model. Fig. 9 shows the results with $X_K$ marked on the curves. 
The accretion rates are the same as used in drawing Fig. 7b above. Note that
in the beginning, the hard X-ray intensity is brightest and the soft component 
is weakest. As the viscosity of the flow is increased and the transition 
radius is reduced, the object becomes softer. This case and that presented 
in Fig. 10b may be suggestive of why the X-ray novae appear brightest first
in the hard X-rays and then become brighter in the soft X-rays. In reality, 
clearly, after the transition radius of around $X_K=10$ is reached, 
the soft X-ray bump continues to rise because of the increase in 
${\dot m}_d$ due to enhanced viscosity. The variation of the spectral 
index in this model is compared with other models in \S 3.4. 

\subsection{When the Sub-Keplerian Component is Absent}

For the sake of argument, we assume that the sub-Keplerian
component does not exist, namely, that the disk is
originally fully Keplerian and subsequently totally 
becomes sub-Keplerian at $X_K$. We show below that we always obtain a soft 
state. This is our Single Component Accretion Flow Model (or, SCAFM). This 
final exercise is to demonstrate the importance of having separate `electron 
fuel' in the form of a sub-Keplerian, inefficiently radiating flow along 
with the Keplerian disk, as assumed in all the models presented above.  
A similar variation as in the previous sub-section could be studied by 
changing $X_K$ as is believed to be induced by viscosity (we call this as 
SCAFM$_2$). The results with ${\dot m}_d=0.01$ are shown in Fig. 10a where 
we mark the respective $X_K$ values identifying each curve. Note that the 
flow is always in the soft state, but since the flow rate is very small, 
the bulk motion Comptonization is negligible as well and the weak hard 
tail of slope $\sim 1.5$ that is seen in soft states cannot be produced 
in this model. This vindicates our claim that the sub-Keplerian component 
is essential to supply an extra set of hot electrons.
In Fig. 10b, we present the spectral index variation with $X_K$ 
(long-dashed curve). For comparison, we plot the same 
quantity for TCAFM3$_2$ (short-dashed curve) and TCAFM2$_2$ 
(solid curve) models. As noted above, SCAFM$_2$ always 
produces a soft state even when the disk rate is very low.
What is interesting, however, is that the spectral index shows a 
distinct minimum. This is because when $X_K$ is very large, the
electron density in the sub-Keplerian region near the transition 
radius $X_K$ is very small and they can be cooled even 
by a small number of soft photons. Similarly, when $X_K$ is
very small, the number of soft photons is very high and can easily cool
the sub-Keplerian flow. Due to the extra sub-Keplerian component 
(${\dot m}_h \ne 0$) in the two component models, the electrons 
are not easily cooled by a few soft photons from the Keplerian
disk. Thus, they tend to produce harder states, except when the Keplerian
disk comes too close to the black hole. In the present Figure, we use
two cases for TCAFM2$_2$ (marked): ${\dot m}_d=0.01$ and $0.001$ for
comparison. For ${\dot m}_d=0.001$ the photons are always too few
to produce soft states. For ${\dot m}_d=0.01$ the soft state 
is produced when $X_K$ is small enough, $\sim 50-60$ (see also, 
e.g., Fig. 6[a-b] for ${\dot m}_d=0.05$). Results of TCAFM3$_2$ is drawn for
${\dot m}_d=0.01$. 

\subsection{Comparison of Model Spectra}

To understand differences among the models we plot the spectra of all the
three models, TCAFM(1-3) in Fig. 11a and the spectral indices
in Fig. 11b. The solid, long-dashed and short-dashed curves are for
TCAFM1, TCAFM2 and TCAFM3 respectively. Three sets of curves are
drawn for Keplerian component rates ${\dot m}_d=0.3,\ 0.05,\ 0.0005$
and the halo rate is fixed at ${\dot m}_h=1$, and $X_K=10$ ($=X_s$ for TCAFM1).
The solution including shocks is the hardest, since for a given set 
of accretion rates, the opacity is highest in the enhanced density 
region when a strong shock is present. This is reflected in the 
spectral index variation as well. For comparison, we present also 
the spectral index variation in the TCAFM1$_2$ model 
where the {\it sum} of the accretion rates is kept fixed.
This is shown as the (lower) short-dash-dotted curve in Fig. 11b. Note that
it has the most pleasant feature of having almost constant spectral
index even when ${\dot m}_d$ is increased by a factor of a thousand.
We have plotted the case where ${\dot m}_{sum}= 
{\dot m}_d + {\dot m}_h = 2.0$ and hence
the spectral index is on the low side ($\sim 0.4-0.5$) in the hard state.
For a smaller ${\dot m}_{sum}=1$ the index is higher (the upper 
short-dash-dotted curve). All the models
definitely bring the system to a soft state much before
the Eddington rate is approached. It is clear that
the Keplerian rate where the soft state is produced depends on 
opacity, which in turn depends on the sum of both the Keplerian 
and the sub-Keplerian accretion rates and the model of the flow, 
i.e., whether the centrifugal barrier is present or not.

As discussed in Paper I, when the accretion rate is very high, the
electrons in the sub-Keplerian region are totally cooled off to a few keV
(see, Fig. 4b above). However, they can still be energized to 
several hundreds of keV by the so-called bulk motion Comptonization
where the relativistically moving electrons directly transfer their
momentum to the trapped soft photons. The resulting spectral index
is $\sim 1.5$ which is observed in the soft states of most of the
black hole candidates (Ebisawa et al. 1996). The variation of the
spectral index with ${\dot m}_d$ (when ${\dot m}_h=1$ is chosen) is
shown in solid curve on the right part of Fig. 11b. Note that there
is a considerable overlap between the Comptonization processes
near ${\dot m}_d \sim 1$. Here, the power-law component should 
be contributed by both the Comptonization processes and thus a break
in power law is expected. This may also have been seen in Cyg X-1
(Zhang et al. 1997). The long-dash-dotted curve drawn in this
region assumes the presence of strong shock (where the density 
goes up by a factor of four for a given accretion rate). As a result,
the slope of $1.5$ is achieved at a much lower disk accretion rate
that that obtained from a spherical convergent flow (solid curve).
Accretion rates obtained from observations probably agree with the
results obtained using this enhanced density region (long-dash-dotted
curve).

\section{COMPARISON WITH OBSERVATIONS OF X-RAY NOVAE}

It is interesting to compare the results of the two component accretion 
flow models presented above with some of the observations of black 
hole candidates. In Fig. 12 spectral evolutions of two well 
known X-ray novae are qualitatively fitted with results of TCAFM1.
The spectral data of GS2000+25 are taken from Tanaka (1991)
and those of GS1124-68 are from Ebisawa (private communication, 1996; 
Ebisawa et al., 1994). For simplicity and to highlight the 
similarity between these two objects, all the 
parameters have been kept fixed (with $X_s=10$
and Schwarzschild black hole of mass $1M_\odot$ which after correction due to
spectral hardening corresponds to $3.6M_\odot$) {\it except} for the rates
${\dot m}_d$ and ${\dot m}_h$. Figs. 12(a-b) show these fits. From the
derived pair of rates (${\dot m}_d$, ${\dot m}_h$) the intermediate
rates are interpolated and the resulting light curves ($1-20$keV) 
are shown in Figs. 12(c-d). Solid and dotted light curves are 
respectively drawn using 
linear-linear and linear-log interpolations of ${\dot m}_d$.
In Fig. 12(c), squared points are obtained from the actual spectra while 
crosses are obtained from the fit in Fig. 12(a). In Fig. 12(d), the squares 
are from ASM light curve of Ginga (Kitamoto, private communication, 1996; Kitamoto 
et al. 1992). The general features of the light curves are clearly
reproduced, including the bumps after $50-70$ days
and another one after $200$ days of outburst. Both show a decay time scale
of $\sim 33$ days. The bump around $50-70$ days shows that the decay of disk accretion
rate is not really exponential, but closer to linear, although after around $200$ days
the disk accretion rate has dropped to the point where the exponential decay would have brought it
anyway. This temporary deviation from exponential decay is suggestive 
of a major change in flow topology, which may occur as the viscosity 
crosses the critical values (C90, C96a). For very low and very high 
viscosities the flow may pass only through the inner sonic point, 
while for intermediate viscosities the flow may have to pass through only
the outer sonic point unless shocks also form 
(see, Fig. 2a of C96a). The rising light 
curves in both the cases were computed in two different ways and 
the results were similar. In one case (using TCAFM1), the ${\dot m}_d$ 
was increased from a quiescent state with an $e$-folding time 
of $\sim 2$ days, while in the other case (using TCAFM3), 
the $X_K$ was reduced exponentially at a similar rate.

The qualitative agreements clearly suggest that a series of quasi-steady
two component solutions can explain the time variation of the 
spectral evolution of the X-ray novae. There is some underestimation of
the hard X-rays for GS 1124-68 nova. This should disappear
when more rigorous fitting routines capable of varying $X_K$ and the
central mass are used.

\section{CONCLUDING REMARKS}

In the present paper, we have made a thorough study of the
spectral properties of two component accretion disk models.
We considered the spectral variation when individual rates are
changed, and also when the location ($X_K$) where the flow
deviates from a  Keplerian  disk is varied. All possible
configurations of the sub-Keplerian flow, including strong and weak shocks, 
no-shocks and the flow without angular momentum are taken into account
as dictated by actual solutions of the governing equations.
Generally speaking, we find that such a model is capable
of explaining most, if not all, of the observational features of
the black hole candidates. Particularly important is the fact that the
models are based on the solid foundation of being constructed out 
of theoretical solutions of black hole accretion, rather than assuming ad hoc
components (as in models with static gaseous [e.g., Haardt \& Maraschi, 
1991] or magnetic [e.g., Galeev, Viana \& Rosner, 1979] corona).
The enhanced density region of the sub-Keplerian flow
outside the horizon adequately serves as the so-called Compton cloud 
which has eluded astrophysicists for the past two decades. The oscillation
of the post-shock region similarly produces quasi-periodic oscillations
as observed in black hole candidates (Molteni, Sponholz \& Chakrabarti, 
1995; Ryu, Chakrabarti \& Molteni, 1996 and references therein). 
The detailed properties of the multiwavelength novae light curves
also come out very naturally from our model.

Whereas a single component, axisymmetric, transonic flow has been tested 
sufficiently well for stability (Chakrabarti \& Molteni, 1995; 
Ryu, Chakrabarti \& Molteni, 1996), the two component models presented 
here have not been adequately tested.
This question is important because two flows of different viscosities 
may give rise to some instabilities at the interface, 
although the effect may be minimized due to the presence of mainly 
{\it supersonic} sub-Keplerian flow above and below the generally sub-sonic 
Keplerian disk. Efforts are being made to perform fully consistent numerical 
simulations, and the results should be reported in the near future.

The author thanks Drs. S. Kitamoto, K. Ebisawa  and L. Titarchuk
for discussions. He is also grateful to the referee, Prof. Paul Wiita 
for reading the manuscript very carefully and suggesting improvements.

{}

\begin{center}
{\bf FIGURE CAPTIONS}
\end{center}
\begin{description}

\item[Fig.~1] Classification of the parameter space spanned by the specific energy
${\cal E}$ and angular momentum $l$ for a thin, inviscid, flow in Schwarzschild geometry.
Regions $SA$ and $SW$ produce standing shocks in accretions and winds respectively.
See text for details.

\item[Fig.~2] Schematic diagram of a multi-component accretion flow. Keplerian disk
(cross-hatched) is flanked by (a generally) quasi-spherical sub-Keplerian halo
which produces a centrifugal pressure supported hot dense region around
the compact object.  In the hard state, $X_{K2}$,
the Keplerian disk becomes sub-Keplerian and produces a giant torus of 
about $10^4R_g$, which collapses as viscosity is increased and the
object goes to soft state. When the shock is absent $X_S\sim X_{K1}$ 
becomes  the centerifugally supported dense region which 
reprocesses soft photons in the same way as the post-shock flow.

\item[Fig.~3] Spectral evolution of an accretion disk with a strong shock
(Model TCAFM1) at $X_s=10$ around a black hole of mass $3.6M_\odot$. The sub-Keplerian halo rate is
${\dot m}_h =1$ and the Keplerian rates are marked on the curves. The dotted curve is
drawn to include the effect of bulk motion Comptonization when ${\dot m}_d=1$.

\item[Fig.~4a-b] (a) Spectral variation and (b) electron temperature variation (dotted curve)
in the TCAFM1$_2$ model where the sum of the disk and the halo rate is kept constant.
In (a), we also include the bulk motion Comptonization when ${\dot m}_d=1.5$ and in (b) we include
the electron Temperature variation in TCAFM1 (Fig. 3) for comparison (solid curve). In TCAFM1$_2$, electron
temperature in hard states remains constant to a greater extent before 
catastrophically becoming too small in the soft state. 

\item[Fig.~5] Spectral variation with the Keplerian disk accretion 
rate ${\dot m}_d$ (as marked) while the halo rate
${\dot m}_h$ is chosen to be unity as before. The result is from the TCAFM3 model
where the sub-Keplerian halo is devoid of angular momentum and therefore
has no centrifugal barrier. Soft states are achieved even when the
accretion disk rate is very low (compare with Fig. 3 and Fig. 4a above).

\item[Fig.~6a-b] (a) Spectral variation of TCAFM3 when $X_K$ is monotonically 
decreased from $400$ to $10$; the curves are drawn for 
$X_K=400,\ 350,\ 300,\ 260,\ 230,\ 200,\ 170,\ 140,\ 120,\ 100,\ 100,\ 90,\ 
50,\ 30,\ 20,$ and $10$ respectively. The corresponding variation in spectral index is 
shown in (b). The greater interception of soft photons causes the black hole 
to go from the hard state to the soft state. Here, the fixed rates are: 
${\dot m}_h=1$,  ${\dot m}_d=0.05$. 

\item[Fig.~7a-b] Similar to Fig. 6a, but $X_K$ (marked on
the curves) could be very far away as in a low accretion rate disk.
${\dot m}_h=1.0$ and ${\dot m}_d=0.001$ in (a)
and $0.01$ in (b). In (a), disk becomes very bright in hard
states before becoming fainter (though remaining hard), while in (b),
disk becomes very bright as $X_K$ is decreased but subsequently
becomes faint as it reached the soft state when $X_K\sim 30$.

\item[Fig.~8]  Spectral variation in TCAFM2. In this case the shock 
is very weak or absent but the centrifugal barrier causes the density
enhancement close to the black hole exactly same as in a shocked flow
(albeit gradually); ${\dot m}_h=1$ and ${\dot m}_d$ are as marked.
Hard to soft state transitions take place due to increase in the 
number of soft photons relative to the electron numbers as in TCAFM1 (Fig. 3).

\item[Fig.~9] Spectral variation in TCAFM2$_2$ model with
$X_K$ marked on the curves. ${\dot m}_h=1.0$ and ${\dot m}_d=
0.01$. For very high $X_K$ the hard X-ray intensity 
is brightest. As viscosity of the flow is increased,
and $X_K$ decreases, the object goes to the soft-state.

\item[Fig.~10a-b] (a) Spectral variation in SCAFM$_2$. Here a single
component Keplerian disk goes completely sub-Keplerian
so that ${\dot m}_h \sim 0$. ${\dot m}_d=0.01$ is chosen. 
$X_K$ values are as marked. Here one always has a soft-component.
Bulk motion Comptonization is neglected. In (b), we 
compare spectral index of this model (long dashed) with those of
TCAFM2$_2$ (solid curves) and TCAFM3$_2$ (dotted curves) as functions of
$X_K$, the location where the flow deviates from a Keplerian disk.

\item[Fig.~11a-b]  Model dependence of spectral properties. 
(a) Solid, long-dashed and short-dashed curves are for TCAFM1, 
TCAFM2 and TCAFM3 respectively. (b) Spectral indices of the 
corresponding models as functions of ${\dot m}_d$. For comparison,
included TCAFM1$_2$ model solutions (short-dash-dotted curves) 
are included, where the sum ${\dot m}$ of the rates is 
kept constant ($2$ for the upper curve and $1$ for the lower curve).
The variation of index due to bulk motion Comptonization is also
included. The region ${\dot m}_d \sim 0.3-1$ may show break in spectral
index as the effects of both the thermal Comptonization and bulk motion 
Comptonization would be present. The long-dash-dotted curve is also
for bulk motion Comptonization but the presence of shock is
assumed which enhances density in the flow for a given accretion rate.

\item[Fig.~12a-d] Qualitative fits of evolution of two X-ray novae 
spectra using two component accretion flow model (TCAFM1) and 
comparison of derived light curves with the observed light curves. 
See text for details.

\end{description}

\clearpage


\clearpage

\begin{thebibliography}{}
\bibitem[ ]{ } Cannizzo, J.K., 1993 in {\it Accretion Disks in Compact Stellar
Systems}, ed. J. Craig Wheeler (Singapore: World Scientific), 6
\bibitem[ ]{ } Chakrabarti, S.K., 1989, ApJ, 347, 365  (C89)
\bibitem[ ]{ } Chakrabarti, S.K. 1990, Theory of Transonic 
Astrophysical Flows (Singapore: World Sci.) (C90)
\bibitem[ ]{ } Chakrabarti, S. K., 1996a, ApJ, 464, 664 (C96a)
\bibitem[ ]{ } Chakrabarti, S.K. 1996b, in Accretion Processes on Black Holes,
Physics Reports, v. 266, No. 5 \& 6, p 229 (C96b)
\bibitem[ ]{ } Chakrabarti, S.K., 1996c, MNRAS, 283, 325
\bibitem[ ]{ } Chakrabarti, S.K., 1997 in  Accretion Phenomena and Related
Outflows, eds. D. Wickramsinghe, L. Ferraro \& G. Bicknell (in press).
\bibitem[ ]{ } Chakrabarti, S.K., \& Molteni, D. 1995,  MNRAS, 272, 80 (CM95) 
\bibitem[ ]{ } Chakrabarti, S.K., \& Titarchuk, L.G., 1995, ApJ, 455, 623 (CT95)
\bibitem[ ]{ } Chakrabarti, S.K., \& Wiita, P.J. 1992, ApJ, 387, L21
\bibitem[ ]{ } Ebisawa, K. et al. 1994, PASJ, 46, 375 (E94)
\bibitem[ ]{ } Ebisawa, K., Titarchuk, L. \& Chakrabarti, S.K. 1996, \pasj, 48, 1
\bibitem[ ]{ }  Galeev, A. A, Rosner, R., \& Viana, G. S. 1979, \apj, 219, 318
\bibitem[ ]{ } Haardt, F. \& Maraschi, L. 1991, \apj, 411, L95
\bibitem[ ]{ } Kazanas, D., Hua, X.M. \& Titarchuk, L.G. \apj (in press)
\bibitem[ ]{ } Kitamoto, S. et al. 1992, \apj, 394, 609
\bibitem[ ]{ } Molteni, D., Sponholz, H. \& Chakrabarti, S.K. 1996 \apj, 457, 805
\bibitem[ ]{ } Paczy\'nski, B., \& Wiita, P.J., 1980, A\&A, 88, 23.
\bibitem[ ]{ } Ryu, D., Chakrabarti, S. K., \& Molteni, D. 1997, \apj (in press).
\bibitem[ ]{ } Shakura, N. I. \& Sunyaev, R. A. 1973, \astap, 24, 337
\bibitem[ ]{ } Shimura, T. \& Takahara, F. 1995, \apj, 445, 780
\bibitem[ ]{ } Sunyaev, R.A. \& Titarchuk, L.G. 1985, A\&A, 143 374 (ST85)
\bibitem[ ]{ } Tanaka, Y. 1991 in Iron Line Diagnostics in X-ray Sources A. 
Treves et al. Springer-Verlag: Heidelberg, 98
\bibitem[ ]{ } Titarchuk, L.G. 1994, ApJ, 434, 570 (T94)
\bibitem[ ]{ } Zhang, S.N. in Accretion Phenomena and Related Outflows,
eds. D. Wickramsinghe, L. Ferraro \& G. Bicknell  (in press).
\end{thebibliography}
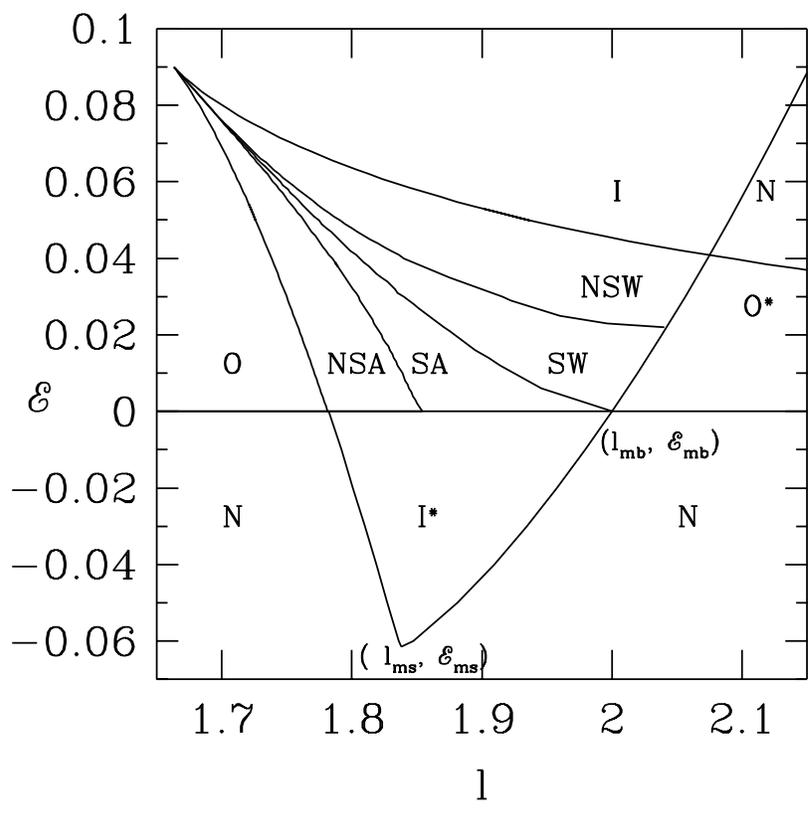
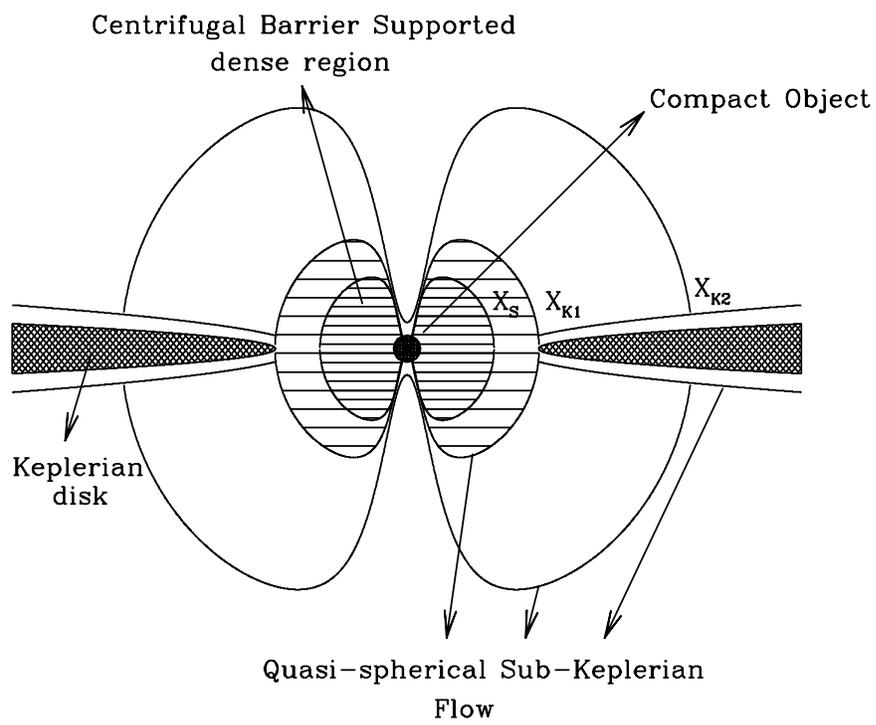
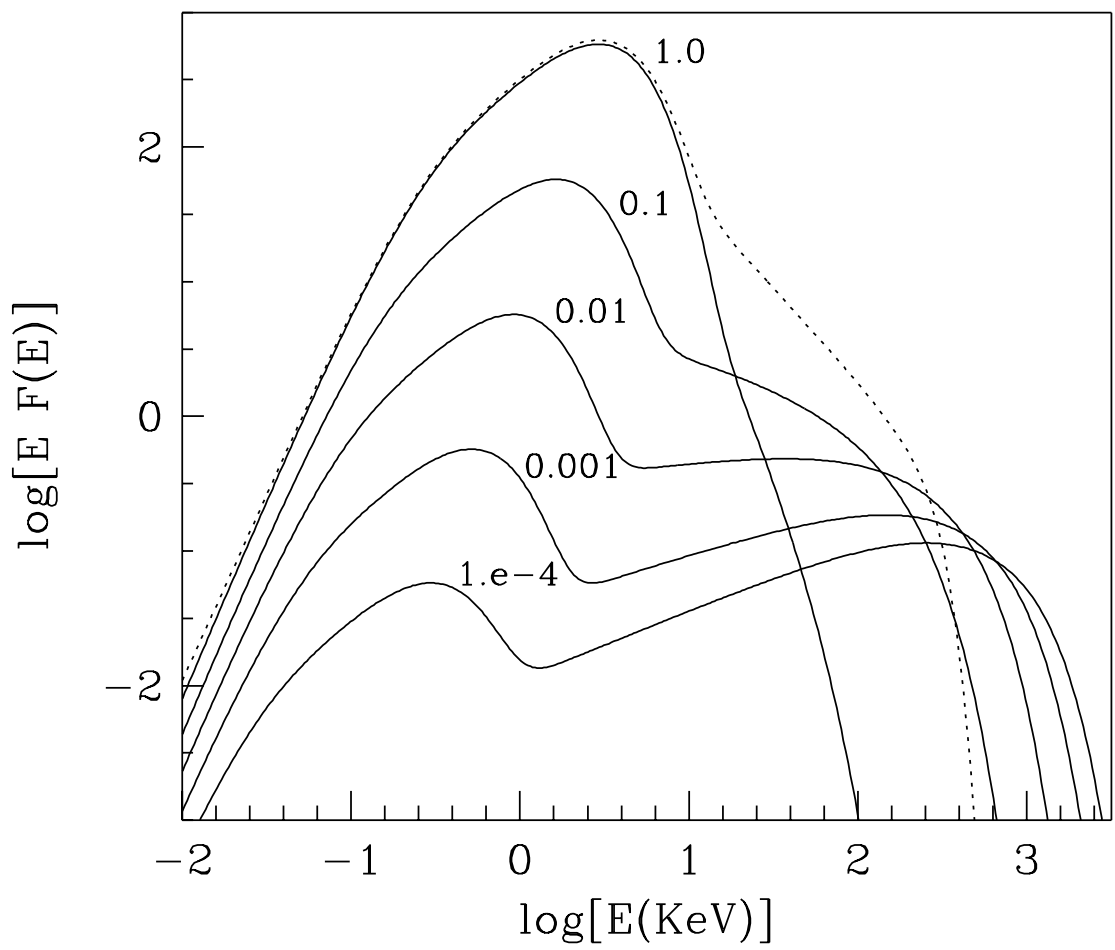
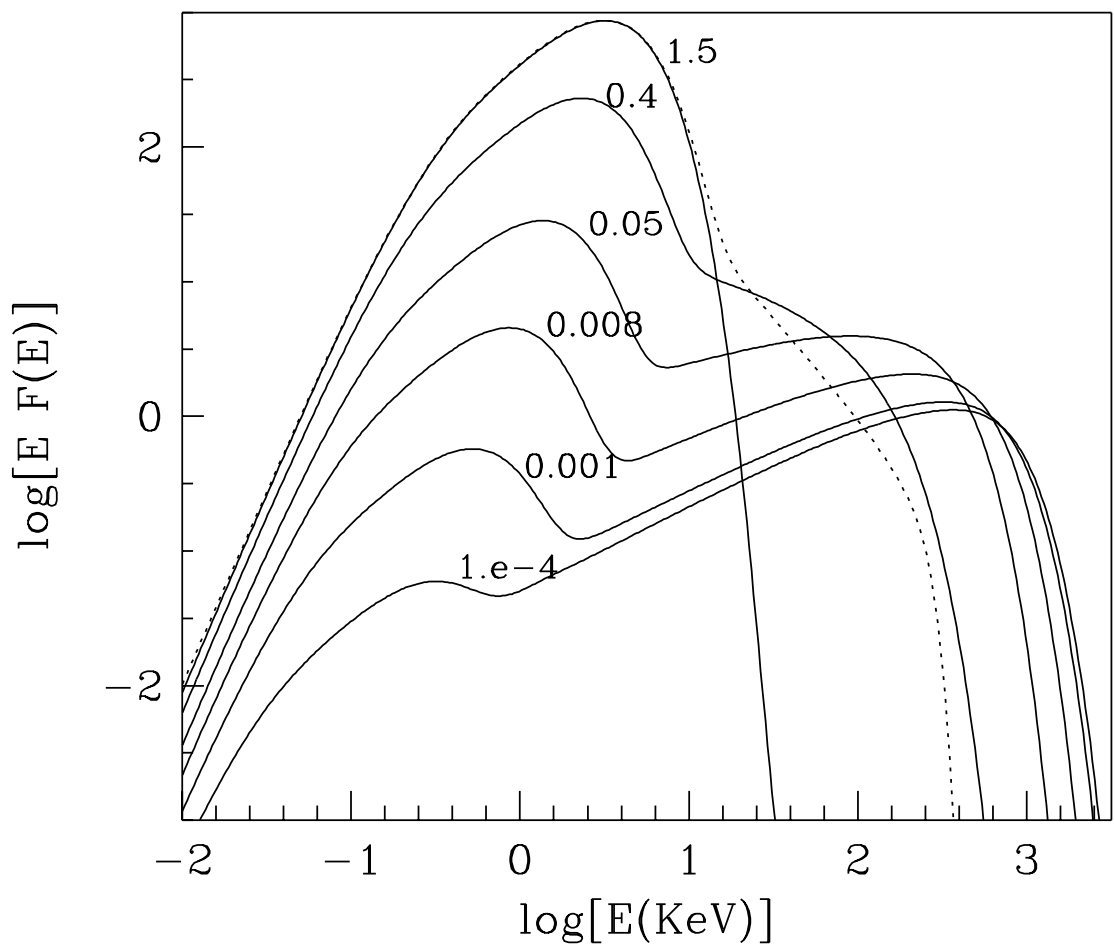
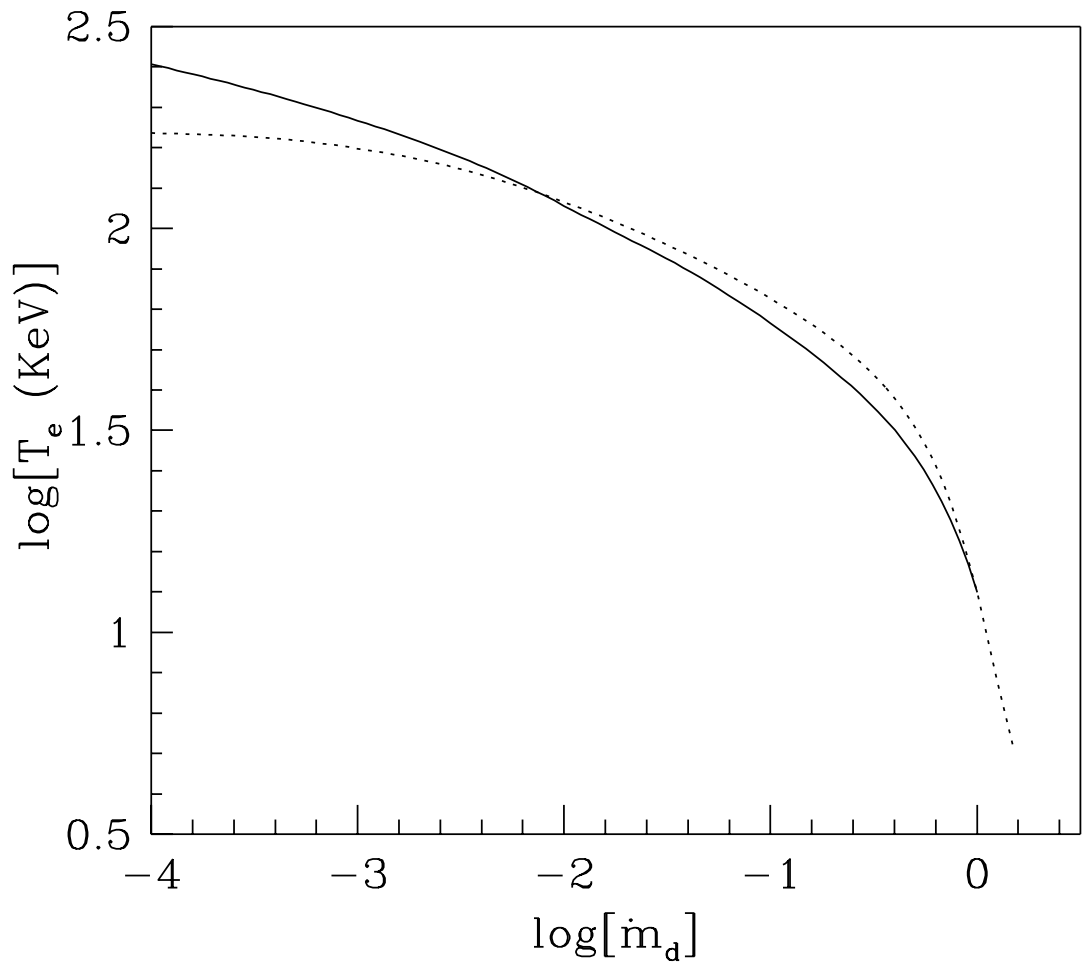
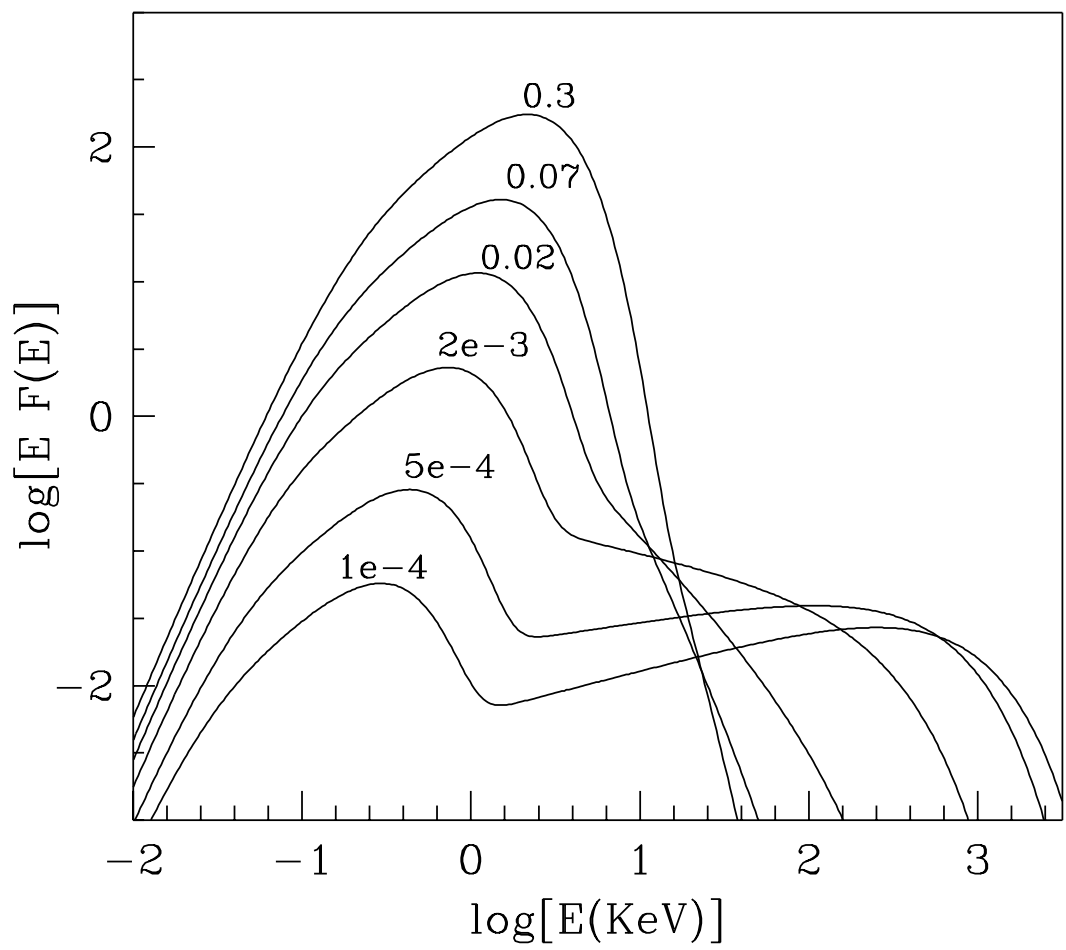
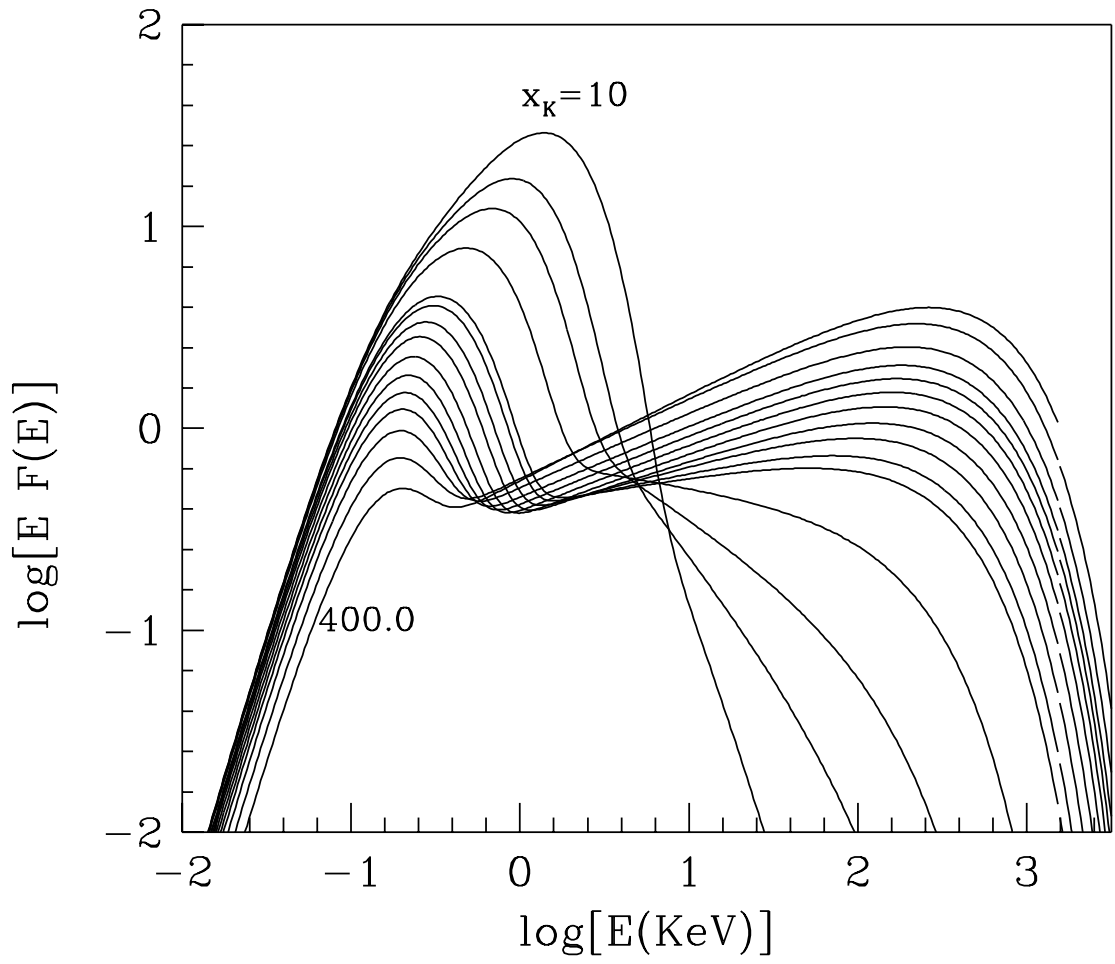
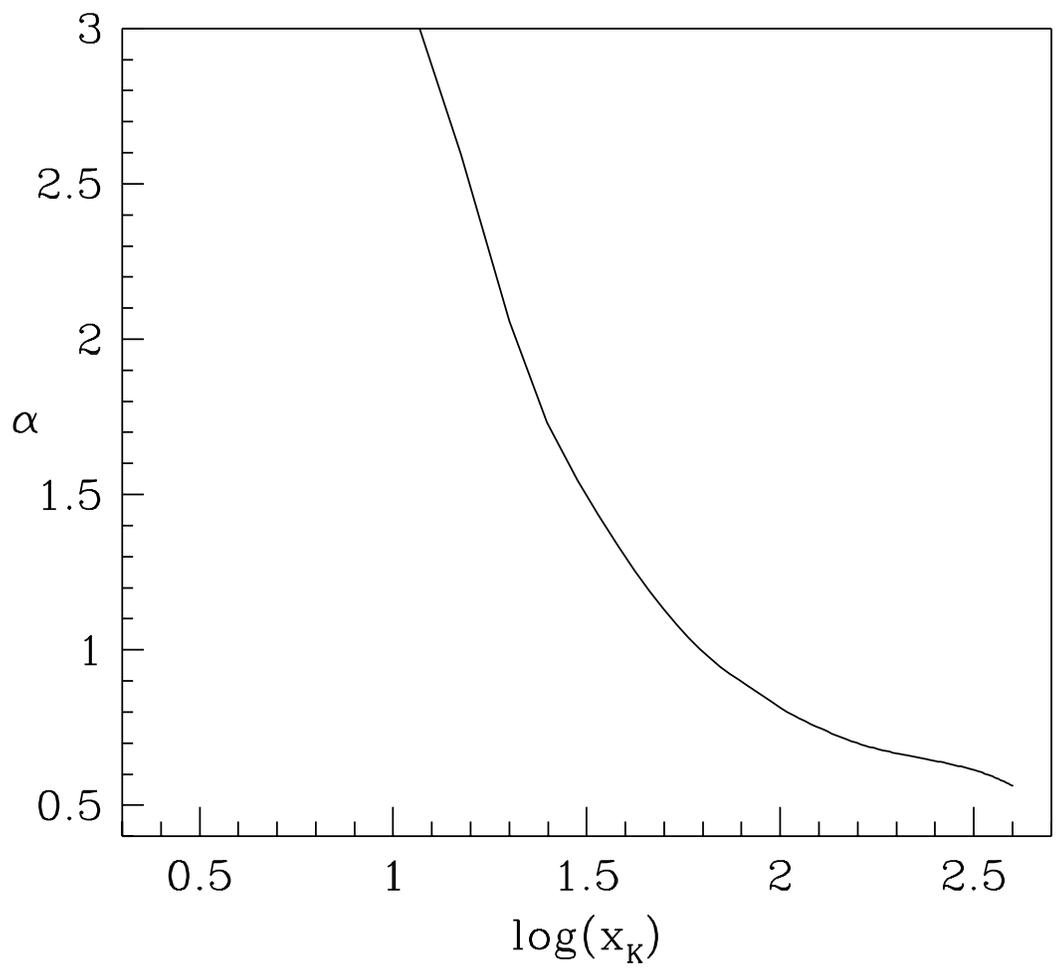
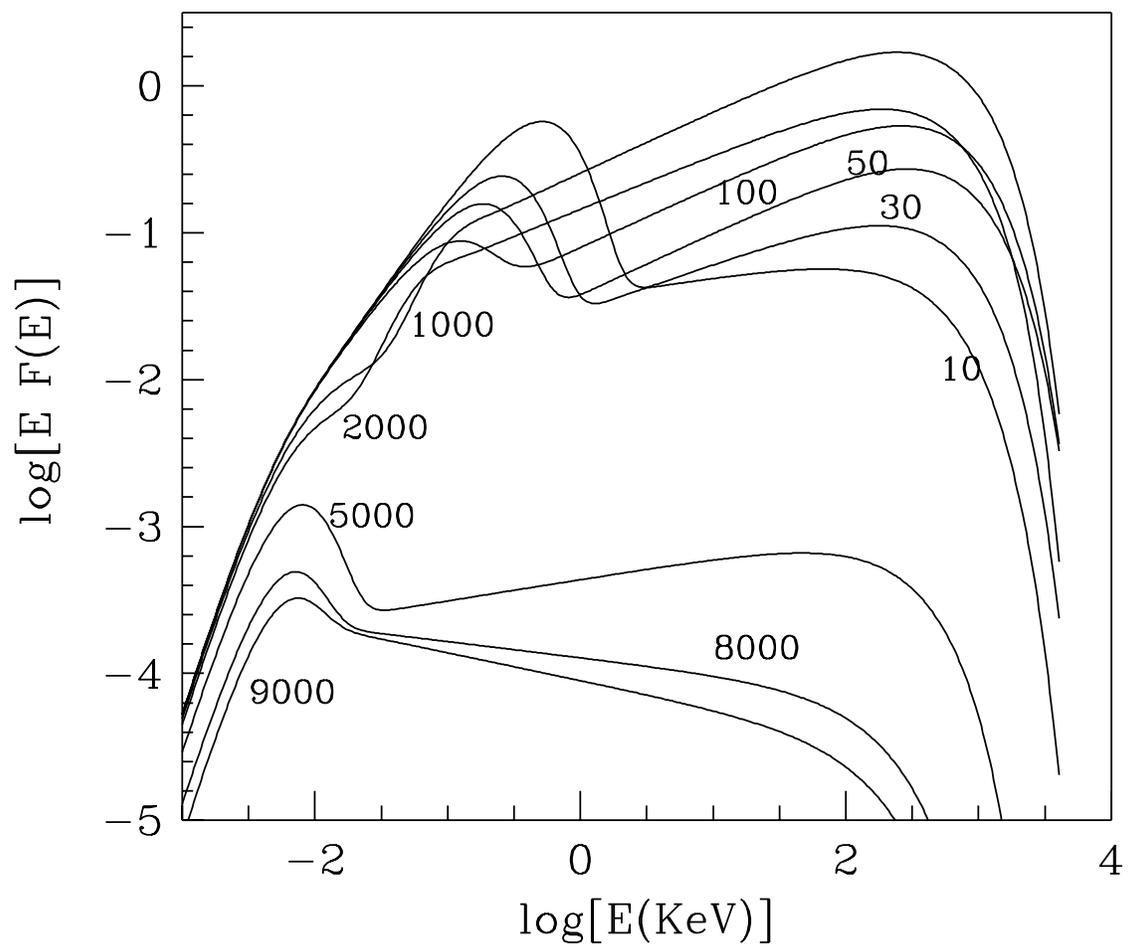
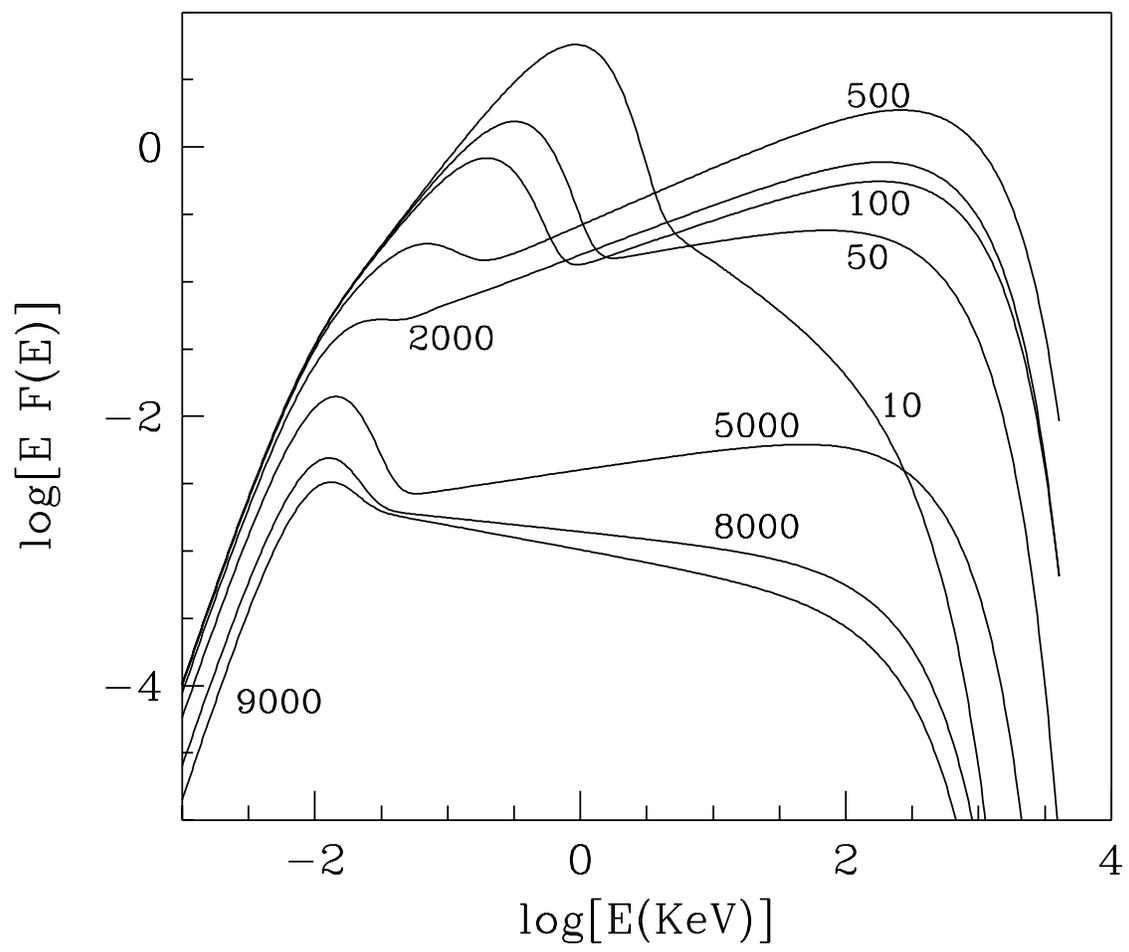
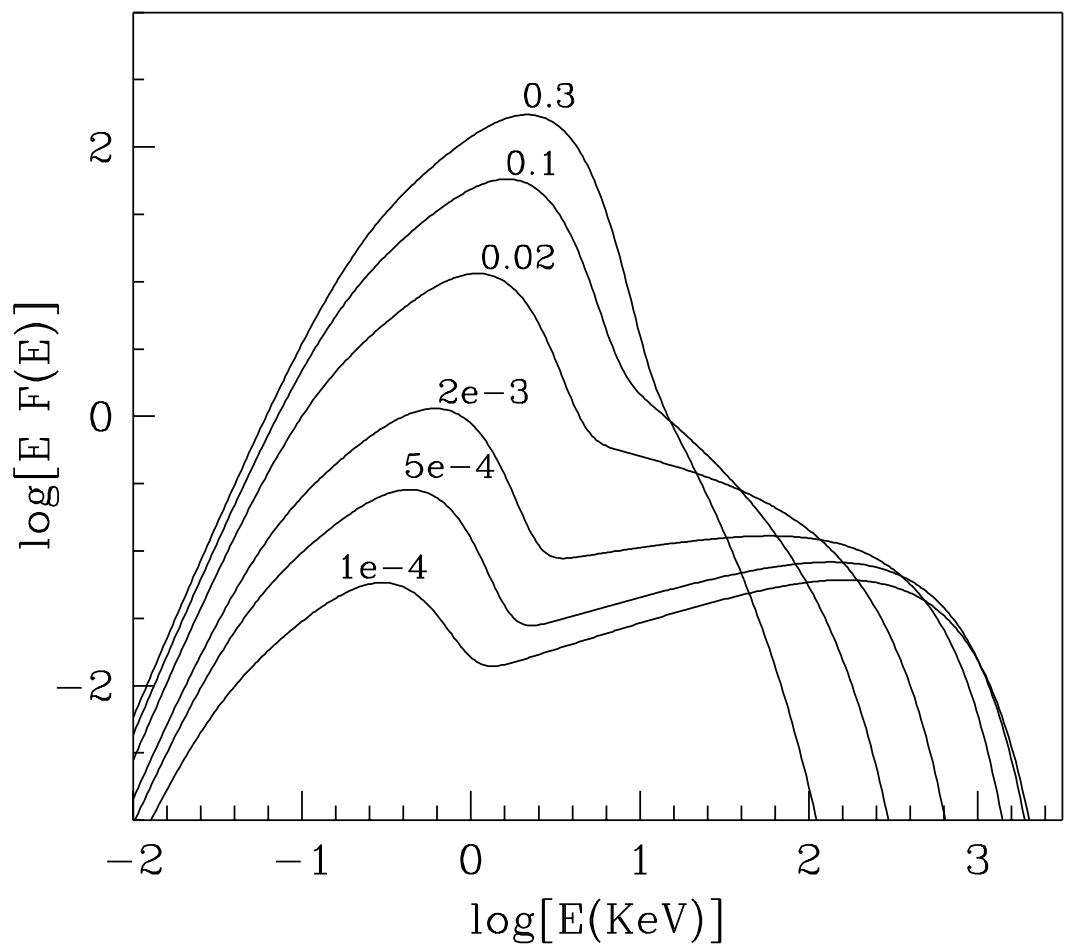
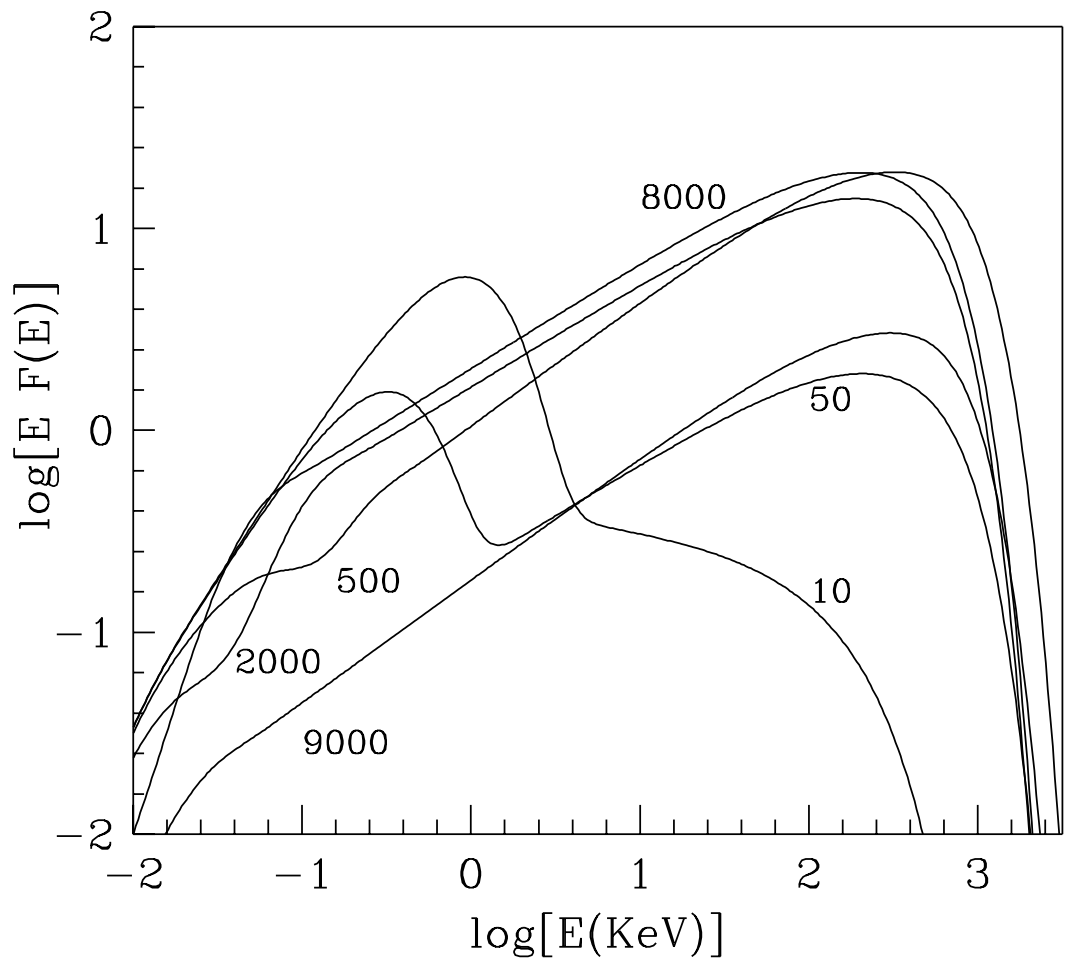
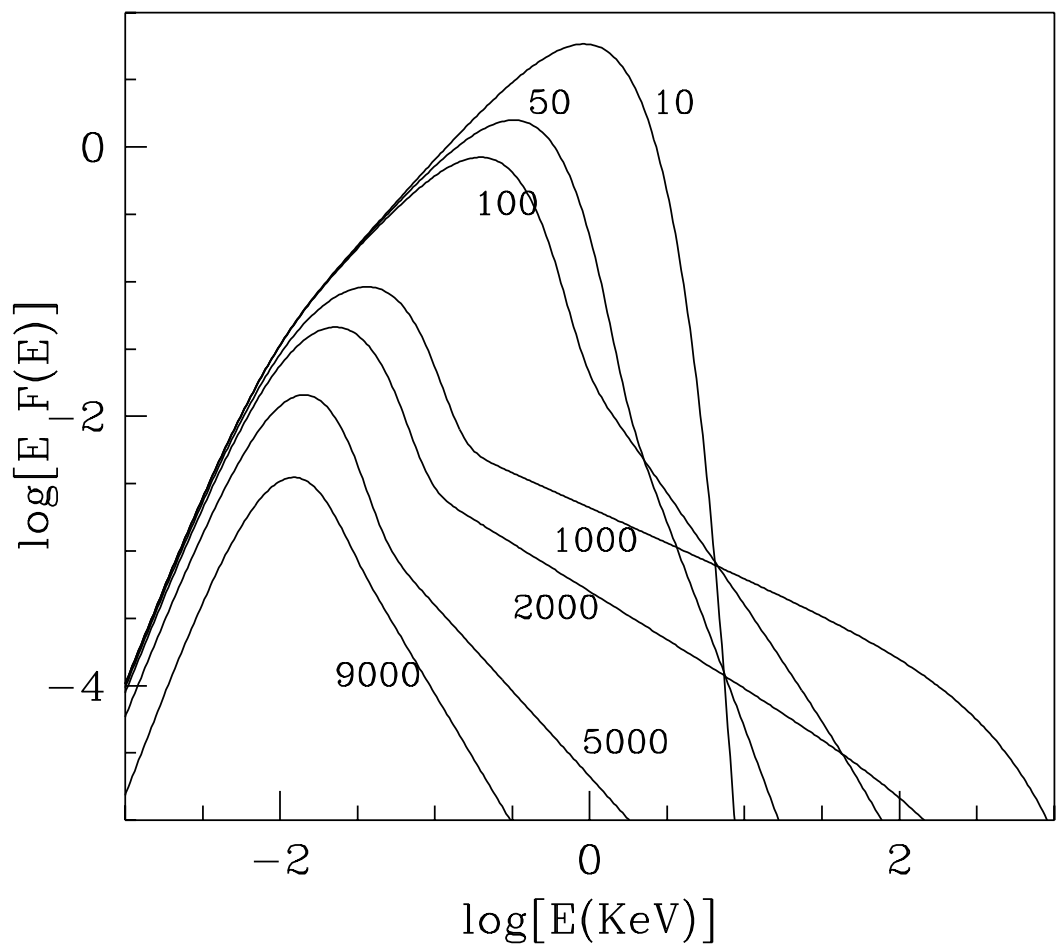
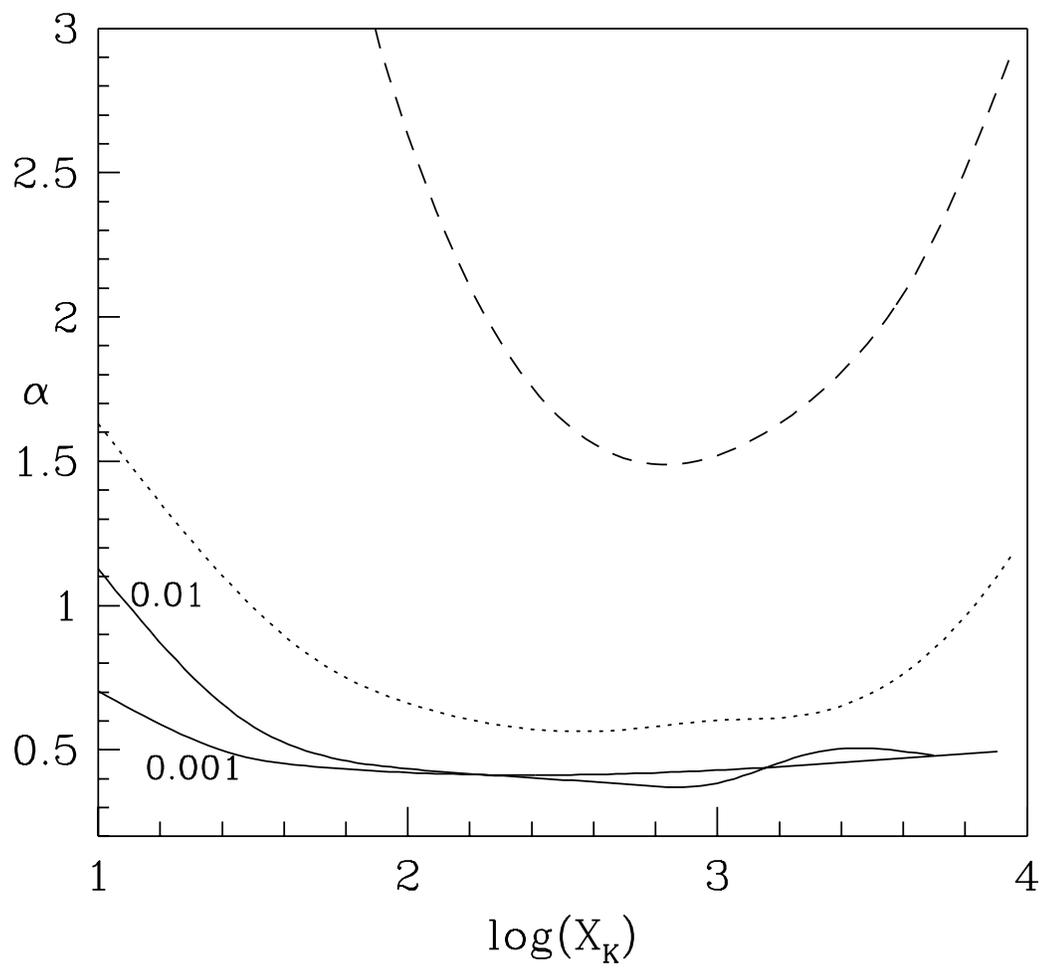
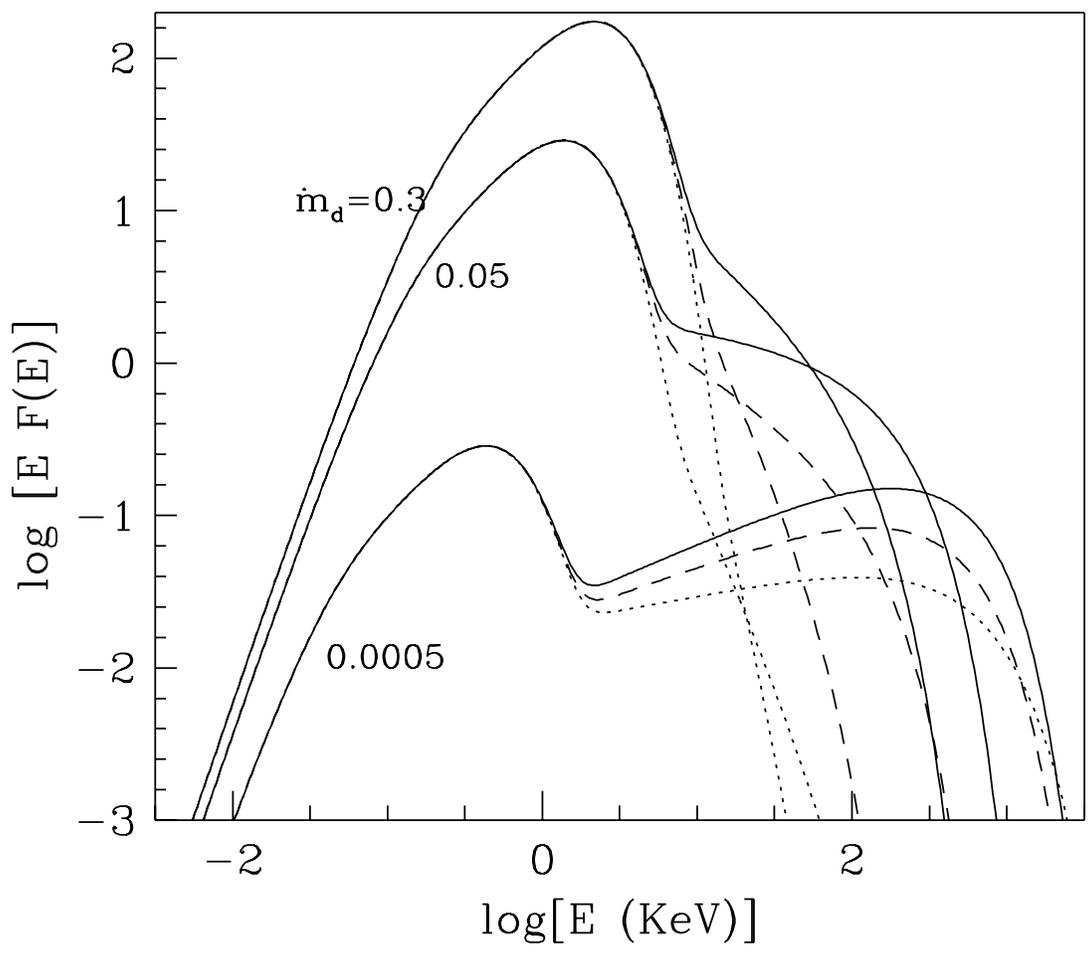
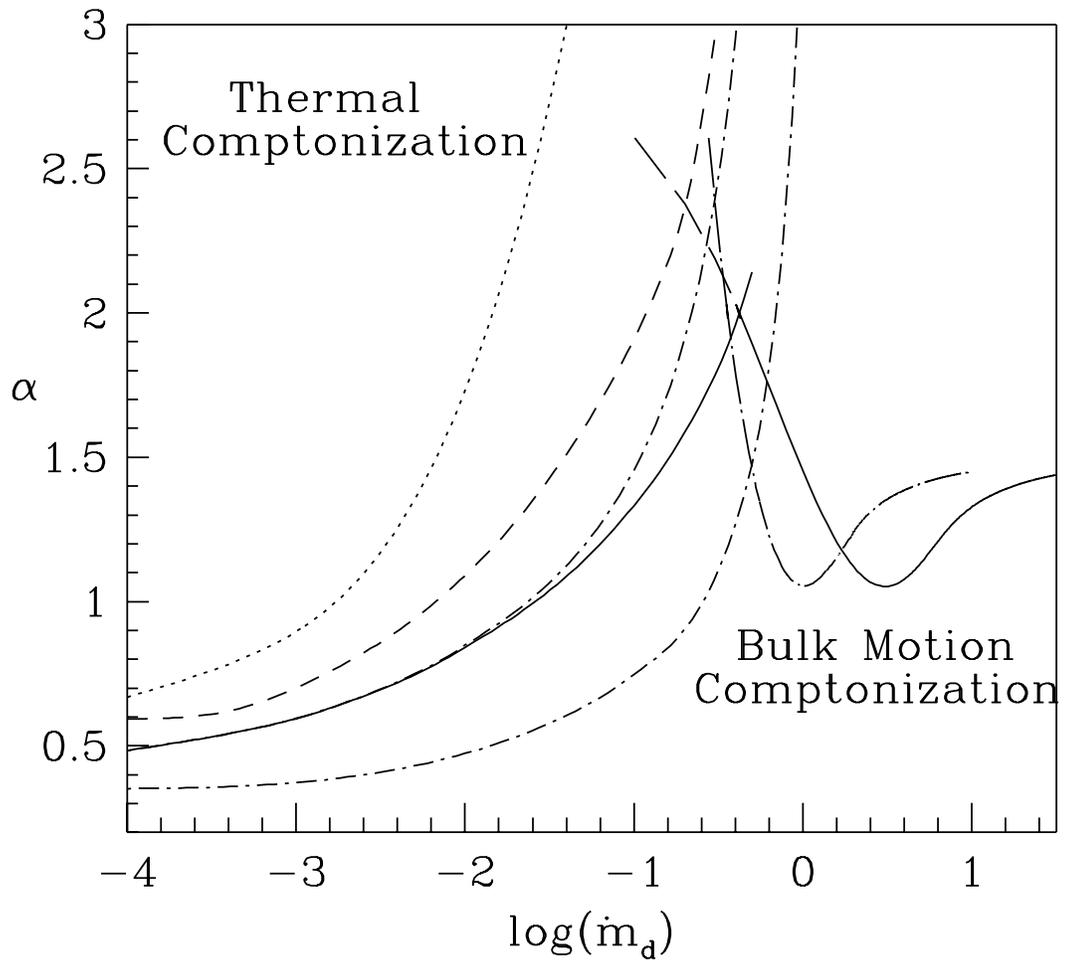
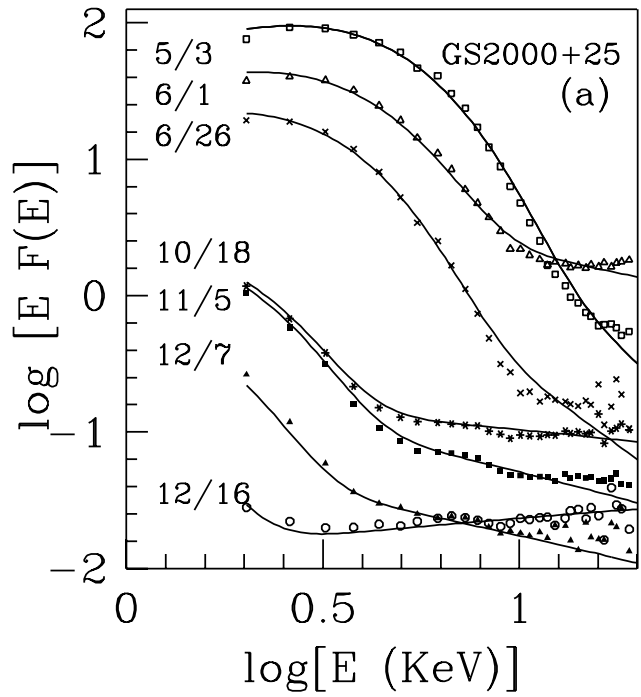
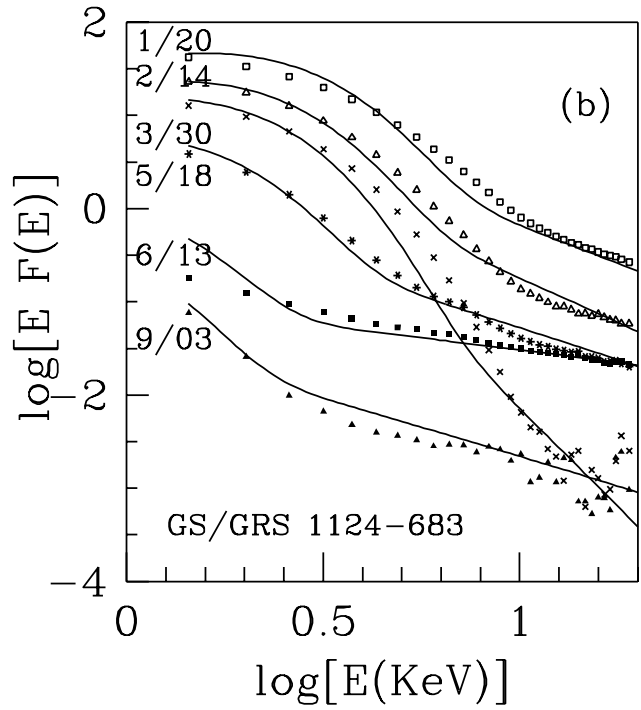
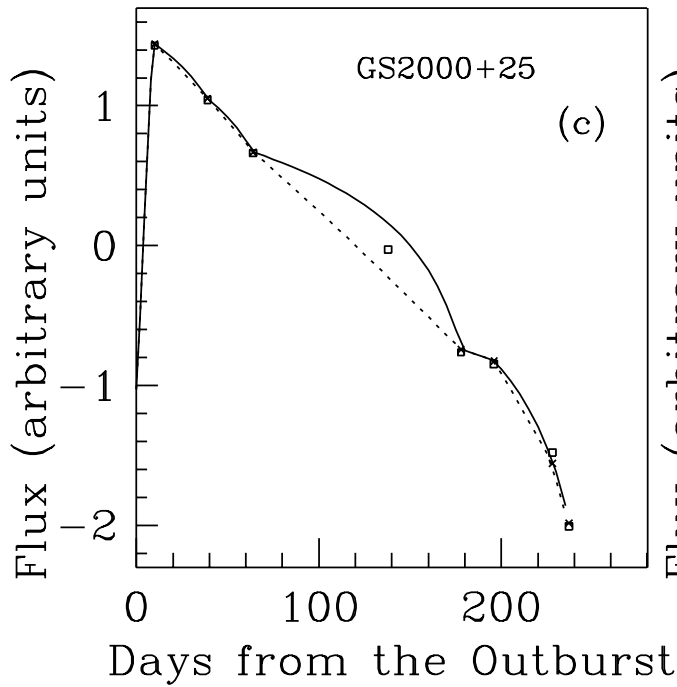
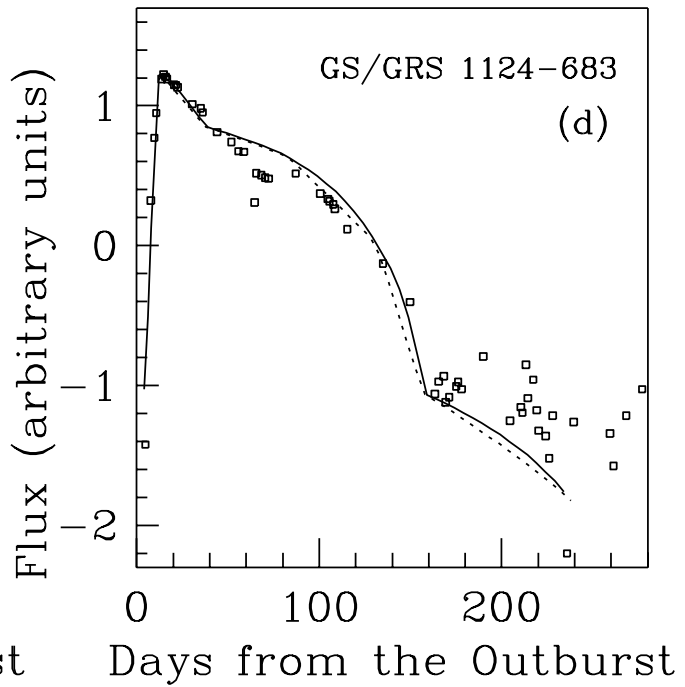
\end{document}